\documentclass[a4paper, amsfonts, amssymb, amsmath, reprint, showkeys, nofootinbib, twoside, onecolumn,notitlepage]{revtex4-1}
\def\prd{{Phys.~Rev.~D}}        % Physical Review D
\def\prl{{Phys.~Rev.~Lett.}}    % Physical Review Letters
\def\nat{{Nature}}              % Nature
\def\physrep{{Phys.~Rep.}}   % Physics Reports
\usepackage{amsmath,amstext}
\usepackage[T1]{fontenc}
\usepackage{amssymb}
\input{epsf}
\usepackage{graphicx}
\usepackage{ae,aecompl}

\usepackage{hyperref}
\usepackage{amsmath}
\usepackage{amssymb}
\usepackage{mathtools}
\usepackage{bm}
\usepackage{cleveref}
\usepackage{tensor}
\usepackage{braket}
\usepackage{enumitem}
\usepackage{mhchem}
\usepackage{amsthm}
\usepackage{physics}

\usepackage{color}

\newcommand{\spc}{\quad \quad \quad}

\def\be{\begin{equation}}
\def\ee{\end{equation}}
\def\beq{\begin{eqnarray}}
\def\eeq{\end{eqnarray}}

\theoremstyle{definition}

\theoremstyle{definition}
\newtheorem{remark}{Remark}

\theoremstyle{theorem}
\newtheorem{theorem}{Theorem}

\begin{document}
\title{Subluminality of relativistic quantum tunneling}
\author{L.~Gavassino, and M.~M.~Disconzi}
\affiliation{Department of Mathematics, Vanderbilt University, Nashville, TN, USA}
\email{lorenzo.gavassino@vanderbilt.edu, marcelo.disconzi@vanderbilt.edu}

\begin{abstract}
We prove that the classical Dirac equation in the presence of an external (non-dynamical) electromagnetic field is a relativistically causal theory. As a corollary, we show that it is impossible to use quantum tunnelling to transmit particles or information faster than light. When an electron tunnels through a barrier, it is bound to remain within its future lightcone. In conclusion, the relativistic quantum tunnelling (if modelled using the Dirac equation) is an entirely subluminal process, and it is not instantaneous.
\end{abstract}

\maketitle

\section{Introduction}

There is some debate over whether the ``speed of tunnelling'' could be faster than the speed of light \cite{Nimtz2002,Buttiker2003,Winful2003,
Buttikersecond2003,
Winful2006,DelBarco2011,DeLeo2013,
Dumont2020,Dumont2022}. Some authors claim that, yes, quantum tunnelling allows for superluminal signalling \cite{Nimtz2002}. Other authors argue that, no, there is no superluminal propagation of particles or signals going on \cite{Buttiker2003}. More recently, some authors \cite{Dumont2020} have proposed an intermediate interpretation: when a particle tunnels through a barrier, it may indeed emerge on the other side faster than light, but the probability for this to happen is so low that in practice photons arrive first, preventing an actual superluminal signalling.
What has made this subject so prone to interpretation is that, if one thinks just in terms of wavepackets and dispersion relations, then it is hard to define unambiguously terms like ``signal'', or ``tunnelling time''. On the other hand, the mathematical theory of partial differential equations provides us with all the tools that are needed for us to settle this matter once and for all. This is what we aim to do here.

\begin{figure}
\begin{center}
\includegraphics[width=0.7\textwidth]{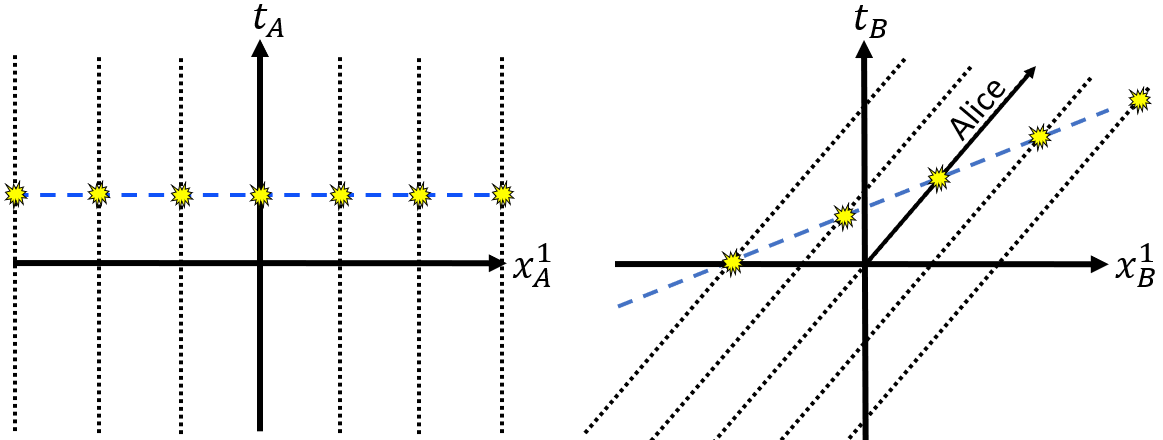}
	\caption{Minkowski diagrams of the thought experiment outlined in the introduction. Left panel: Alice's viewpoint. The lightbulbs (black dashed lines) are at rest, and they are all turned on simultaneously at $t_A=1$ (blue dashed line). In Alice's coordinates, the spacetime event at which the $k$-th bulb is turned on is $(1,k)$, and it is marked with a yellow star. Right panel: Bob's viewpoint. The event at which the $k$-th bulb is turned on is now $(\gamma+\gamma v k, \gamma v +\gamma k)$. These events are no longer simultaneous for different values of $k$, and it looks as if there was a ``signal'' travelling along the blue dashed line $x_B^1=t_B/v -(\gamma v)^{-1}$.}
	\label{fig:aliceandbob}
	\end{center}
\end{figure}

Let us first clarify what is mean by ``superluminality'' in this context. Consider the following thought experiment. In the reference frame of Alice, there is a sequence of lightbulbs at rest, at a distance of one meter from each other. Alice has synchronized them in such a way that they all turn on simultaneously at $t_A=1$, according to Alice's clock (see figure \ref{fig:aliceandbob}, left panel). Now, let us move to Bob's frame, who travels with velocity $-v$ with respect to Alice. By relativity of simultaneity \cite{special_in_gen,GavassinoUnstableParticles,GavassinoSuperluminal2021}, the bulbs  do not turn on all together, according to Bob. Instead, they turn on in sequence, and it looks as if there were a ``superluminal impulse'' travelling at speed $v^{-1}>1$ (we set $c=1$), which commands the bulbs to turn on one after the other. Clearly, this illusory ``impulse'' is just an artefact of synchronization. A similar phenomenon occurs whenever the phase velocity of a wave (or the group velocity of a wavepacket) is larger than the speed of light while the underlying theory is causal \cite{Susskind1969}: different regions of the system are synchronized to generate what looks to be a ``superluminal wave'', but no actual transfer of information or energy occurs \cite{TolmanTheoryOfRelativity1917}.

The ``apparent superluminality'' discussed above is \textit{not} what we are interested in. Also, we are not interested in issues related to the ontology of the wavefunction, which may render all quantum mechanics superluminal at the outset. Consider the following example. An electron is in a quantum superposition, with $1/2$ probability of being on Earth, and $1/2$ probability of being in the Andromeda Galaxy. If we make a measurement, and we detect the electron, then automatically we know that it is not in the Andromeda Galaxy. However, was the electron already on Earth, say, one second before the measurement, or did our measurement itself ``localise'' the electron on Earth? Could it be that, maybe, the electron was on the Andromeda Galaxy one second before the measurement, and then it ``teleported itself'' on Earth at the instant of the measurement? The answer depends on the interpretation of quantum mechanics one is adopting, and it is intrinsically unobservable. For this reason, we will leave this kind of problem aside.

The question that we aim to answer here is more practical: taking into account the statistical nature of quantum mechanics, is it possible to \textit{use} quantum tunnelling as a means to effectively transfer particles or signals faster than light? To make this question more precise, we have identified three rigorous ``practical'' notions of superluminality, which will be assessed here, one by one:
\begin{itemize}
    \item Can the support of the wavefunction propagate outside the lightcone? If this could happen, it would be possible to make two consecutive measurements, where the electron is first detected inside some region $\mathcal{R}$, and then outside the causal future of $\mathcal{R}$. This would mean that we can actually \textit{observe} an electron making a superluminal jump. It is well-known that, in the absence of a potential barrier, such process is forbidden within relativistic quantum mechanics   \cite{Thaller1992}. In section \ref{implicazia}, point (i), we will prove that the same is true also in the presence of a barrier.
    \item Can Alice use quantum tunnelling to send a message to Bob, assuming that Bob sits outside the future lightcone of Alice? There is general consensus that this should not happen, as it would constitute a violation of the principle of causality. However, at present there is no rigorous mathematical proof of this fundamental impossibility within tunnelling models. We will provide such proof in section \ref{implicazia}, point (ii).
    \item Suppose that the electron is on one side of the barrier, with probability $\mathcal{P}$. Can the probability of detecting the electron on the other side become larger than $1-\mathcal{P}$ in less time than it would take for light to travel between the two edges of the barrier? If this were possible, we would be able to use the barrier to ``transfer'' probability (and, thus, particles) faster than light. In section \ref{powerfuliequal}, we prove that this eventuality is indeed forbidden: probability ``flows''  subluminally between the edges of the barrier. As a corollary, quantum tunnelling is not instantaneous.
\end{itemize}

Throughout the article, we adopt the spacetime signature $(-,+,+,+)$, and work in natural units: $c=\hbar=1$. We use standard rectangular coordinates $\{x^\alpha\}_{\alpha=0}^3$ in Minkowski space $\mathbb{R}^{1+3}$, with $t:=x^0$ denoting a time coordinate. Greek indices vary 
from $0$ to $3$, Latin indices from $1$ to $3$, and the sum convention is adopted.

\section{A simple theorem}

Our goal is to assess whether previous claims of superluminal physics (e.g. \cite{DeLeo2013,
Dumont2020}) are mathematically rigorous. Since all such claims are derived within the framework of relativistic quantum mechanics, we will also stick to this approach (although the final word on the subject should come from quantum field theory \cite{Peskin_book}). In particular, following \citet{Dumont2020}, we will consider a single electron with quantum dynamics governed by the classical Dirac equation (we adopt the sign conventions of \citet{Weinberg_book_1972}):
\begin{equation}\label{diracequaz}
(\gamma^\mu \partial_\mu + ie\gamma^\mu A_\mu +m)\Psi =0 \, .
\end{equation}
Here, $\gamma^\mu$ are Dirac's gamma matrices, $\Psi=\Psi(x)$, $x\in \mathbb{R}^{1+3}$, is a classical Dirac spinor (representing the electron \cite{landau4}), $e$ and $m$ are the electron's charge and mass. The field $A_\mu=A_\mu(x)$, $x\in \mathbb{R}^{1+3}$, is the electromagnetic four-potential, and it is treated as a fixed, assigned, smooth function of the coordinates (it is not a dynamical degree of freedom). For tunnelling models, one should take\footnote{One should be careful about the sign: the potential energy of a particle with charge $q$ in an electrostatic potential $\phi$ is $V=q\phi$. For the electron, $q=-e$. Furthermore, given that our metric signature is $(-,+,+,+)$, we have that $\phi=A^0=-A_0$. Thus, $V=q\phi=eA_0$.}
\begin{equation}\label{onlyelectric}
eA_\mu =(V,0,0,0) \, ,
\end{equation}
where $V(x)$ is the potential energy barrier. However, here we may also keep the potential $A_\mu$ completely general. 

Our first task is to compute the characteristics of the system \cite{CourantHilbert2_book}. As a system of first-order partial differential equations, the Dirac equation \eqref{diracequaz} is naturally written in the standard matrix form (recall that $\Psi$ is has four components):
\begin{equation}\label{represento}
\mathcal{M}^\mu \partial_\mu \Psi + \mathcal{N} \Psi =0 \, ,
\end{equation}
where $\mathcal{M}^\mu=\gamma^\mu$ and $\mathcal{N}=ie\gamma^\mu A_\mu +m$ are $4 \times 4$ complex matrices. Working in the Weyl basis, we can write $\mathcal{M}^\mu$ explicitly  \cite{weinbergQFT_1995}:
\begin{equation}\label{gammasss}
\mathcal{M}^0 =\gamma^0 =-i
\begin{bmatrix}
   0_{2\times 2} &  \mathbb{I}_{2 \times 2}  \\
    \mathbb{I}_{2 \times 2}  & 0_{2\times 2}   \\
\end{bmatrix} \,  \quad \quad \quad
\mathcal{M}^j= \gamma^j =-i
\begin{bmatrix}
   0_{2\times 2} &  \sigma_j  \\
    -\sigma_j  & 0_{2\times 2}   \\
\end{bmatrix} \, ,
\end{equation}
where $\sigma_j$ are the Pauli matrices:
\begin{equation}
\sigma_1 =
\begin{bmatrix}
   0 &  1  \\
    1  & 0   \\
\end{bmatrix} \,  \quad \quad
\sigma_2 =
\begin{bmatrix}
   0 &  -i  \\
    i  & 0   \\
\end{bmatrix} \,  \quad \quad
\sigma_3 =
\begin{bmatrix}
   1 &  0  \\
    0  & -1   \\
\end{bmatrix} \, .
\end{equation}
The characteristic surfaces are defined as the surfaces $ \Phi=\text{const} $ of any scalar field $\Phi$ such that
$
\det[\mathcal{M}^\mu \xi_\mu] = 0 
$,
with $\xi_\mu = \partial_\mu \Phi$. Using equation \eqref{gammasss}, we obtain
\begin{equation}\label{dterminant}
0=\det[\mathcal{M}^\mu \xi_\mu] = (-i)^4 \det 
\begin{bmatrix}
   0 &  0 & \xi_0 +\xi_3 & \xi_1 - i \xi_2  \\
   0 &  0 & \xi_1 + i\xi_2 & \xi_0 - \xi_3  \\
   \xi_0 -\xi_3 &  -\xi_1 + i\xi_2 & 0 & 0  \\
   -\xi_1 -i\xi_2 &  \xi_0+\xi_3 & 0 & 0  \\
\end{bmatrix} = (\xi_\mu \xi^\mu)^2 \, .
\end{equation}
Hence, $\xi_\mu = \partial_\mu \Phi$ must be lightlike, meaning that the characteristic surfaces are null surfaces. This immediately allows us to derive the following
\medskip

\begin{theorem}[Causality of the Dirac equation]\label{theo}
Assume that $A_\mu$ is continuously differentiable and let $\Psi$ be a continuously
differentiable solution to \eqref{diracequaz}.
Let $\Sigma \subset \mathbb{R}^{1+3}$ be a Cauchy surface. Then, for any point $x$ in the future of $\Sigma$, the value of $\Psi$ at $x$, i.e., $\Psi(x)$, depends only on the 
values of $\Psi$ in the region $J^-(x) \cap \Sigma$, and on the value of $A_\mu$ in the region $J^-(x) \cap J^+(\Sigma)$. Here, $J^-(x)$ is the causal past of $x$, and $J^+(\Sigma)$ is the causal future of $\Sigma$. 
\end{theorem}
\begin{remark}
In practice, one usually takes $\Sigma$ to be a surface where initial data 
for the system \eqref{diracequaz} is given (e.g., $\Sigma = \{ t = 0\}$).
In this case, the conclusion of the theorem can be rephrased in a more intuitive
form as saying that the value of $\Psi$ at $x$, i.e., $\Psi(x)$, depends only on the 
initial data in the region $J^-(x) \cap \Sigma$, and on the value of $A_\mu$ in the region $J^-(x) \cap J^+(\Sigma)$.
\end{remark}

\begin{proof}
Fix $x$ in the future of $\Sigma$ and let
$\Psi_1$ and $\Psi_2$ be two continuously differentiable solutions  of the Dirac equation corresponding to two different choices of external potential. Then we have $\mathcal{M}^\mu \partial_\mu \Psi_1 + \mathcal{N}_1 \Psi_1 =0$ and $\mathcal{M}^\mu \partial_\mu \Psi_2 + \mathcal{N}_2 \Psi_2 =0$. Now assume that the external potential is the same on the spacetime region $J^-(x) \cap J^+(\Sigma)$. Then, if we restrict our attention to such region, we have $\mathcal{N}_1=\mathcal{N}_2$, and the field $\Psi_{\text{diff}}:=\Psi_1-\Psi_2$ is a solution of $\mathcal{M}^\mu \partial_\mu \Psi_{\text{diff}} + \mathcal{N}_1 \Psi_{\text{diff}} =0$ on $J^-(x) \cap J^+(\Sigma)$. Finally, assume that 
$\Psi_1$ and $\Psi_2$ agree on $J^-(x) \cap \Sigma$. Then $\Psi_{\text{diff}}=0$ on $J^-(x) \cap \Sigma$. At this point, we can just apply John's Global Holmgren Theorem (see \citet{Rauch_book}, Section 1.8), considering that the characteristics of the Dirac equation are the same as those of the wave equation, and we find that $\Psi_{\text{diff}}=0$ on $J^-(x) \cap J^+(\Sigma)$. This implies $\Psi_1(x)=\Psi_2(x)$. 
\end{proof}

We observe that the conclusion of Theorem \ref{theo} is coordinate
independent, even if we employed standard coordinates in the computation of the 
characteristics. This follows from the invariance of the characteristics (see, e.g.,
\cite{CourantHilbert2_book}) and of $J^\pm$ \cite{Hawking1973}, as well as from standard theory of hyperbolic differential equations \cite{LerayBook-1953,HormanderBook-AnalysisPD-2003-3}. We also remark that Theorem \ref{theo} is not new. The Dirac equation
is known to be a hyperbolic partial differential equation (see, e.g., \cite{Choquet-BruhatBook-2009}) and thus 
Theorem \ref{theo} follows from textbook theory (above we quoted \citet{Rauch_book}
in order to provide the reader with a precise reference, but there are plenty
of sources explaining the properties we used for the proof, e.g.,
\cite{LerayBook-1953,HormanderBook-AnalysisPD-2003-3,Choquet-BruhatBook-2009,LichnerowiczBookMHD-1967,JohnBook-1982,KajitaniKajitaniBook-1991,MizohataBook-1985,TrevesBook-1975,SobolevBook-1991,Taylor-1996-1}; see
the appendix of \cite{DisconziExistenceCausalityConformal} for a summary). Nevertheless,
we felt the need to state Theorem \ref{theo} and provide its proof because,
as the literature review presented in the introduction demonstrates, there seems
to be some confusion in the literature regarding the causal properties of the 
Dirac equation. In particular, properties that follow from standard hyperbolic
theory seem to be neglected in these discussions.

\begin{figure}
\begin{center}
\includegraphics[width=0.5\textwidth]{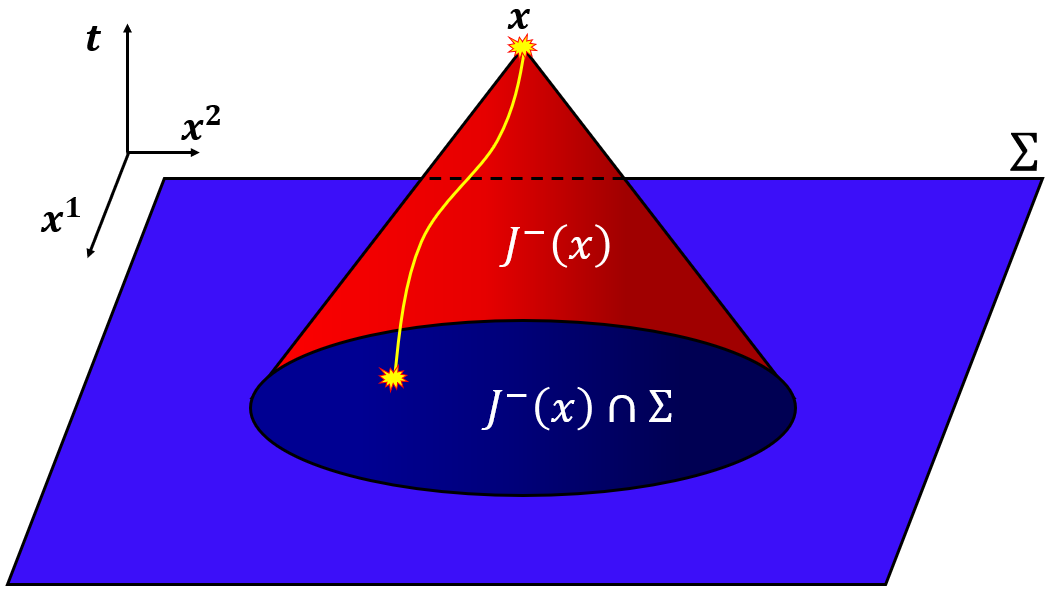}
	\caption{Relativistic principle of causality. Let $\Sigma$ (blue plane) be the initial-data hypersurface, e.g. the hyperplane $\{t=0 \}$. Pick an event $x$ in the future of $\Sigma$. Such event can be influenced only by that portion of $\Sigma$ that can be reached by a non-spacelike worldline emitted from $x$ (yellow line). In other words, the value of $\Psi(x)$ depends only on the initial data prescribed \textit{inside} the past light-cone of $x$ (in red). Changes of initial data outside the past lightcone of $x$ cannot affect the value of $\Psi(x)$, when we solve the Dirac equation. Furthermore, we cannot change the value of $\Psi(x)$ even by altering the external potential $A_\mu$ outside the past lightcone of $x$.}
	\label{fig:fig1}
	\end{center}
\end{figure}

Theorem \ref{theo} coincides with the principle of relativistic causality that we meet in all textbooks of General Relativity \cite{Hawking1973,Wald,carroll_2019}, and in the literature of relativistic hydrodynamics \cite{BemficaCausality2018,Bemfica2019_conformal1,
Causality_bulk,BemficaPRL2021,
BemficaDNDefinitivo2020,
GavassinoSuperluminal2021,GavassinoCausality2021,GavassinoStabilityCarter2022}. It is the mathematical condition that people have in mind when they say: ``no signal can exit the lightcone'' \cite{Susskind1969} (see figure \ref{fig:fig1}).

Let us make an interesting remark. If we set $A_\mu=0$, then we recover the free Dirac equation. It is well known that, in this case, $\Psi$ is also a solution of the free Klein-Gordon equation. Therefore, it is quite trivial to see that the free Dirac equation is a relativistically causal equation \cite{Fox1970}. However, when $A_\mu \neq 0$ (e.g. inside a potential barrier), this becomes less intuitive. The key insight, here, is that the propagation of information is \textit{entirely} determined by the characteristics of the system, which depend only on the principal part of the Dirac equation (the part with highest derivatives: $\gamma^\mu \partial_\mu \Psi$) and are completely unaffected by the presence of the external potential $A_\mu$. In a nutshell, the presence of a potential barrier cannot increase (or shorten) the ``speed of information''.

\section{Implications}\label{implicazia}

\begin{figure}
\begin{center}
\includegraphics[width=0.7\textwidth]{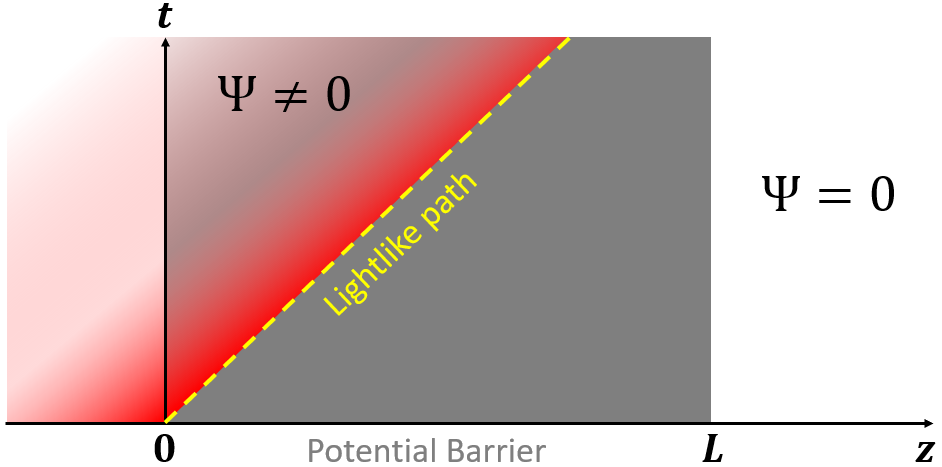}
\includegraphics[width=0.7\textwidth]{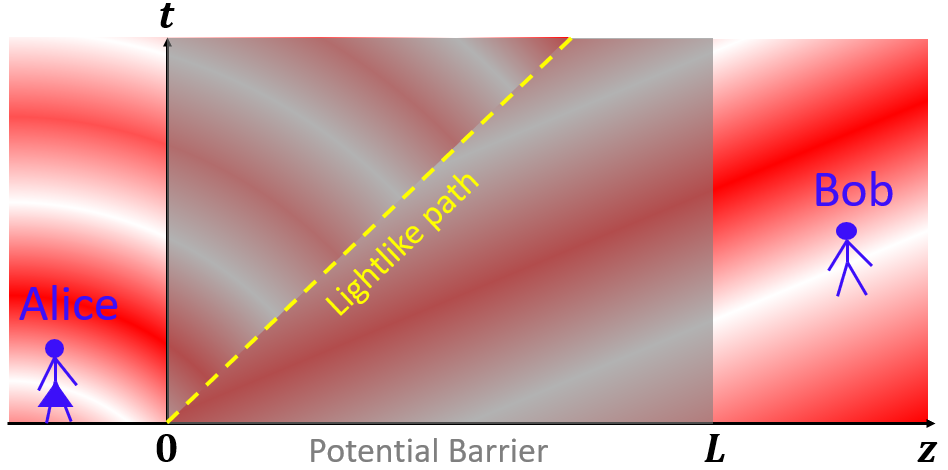}
\includegraphics[width=0.7\textwidth]{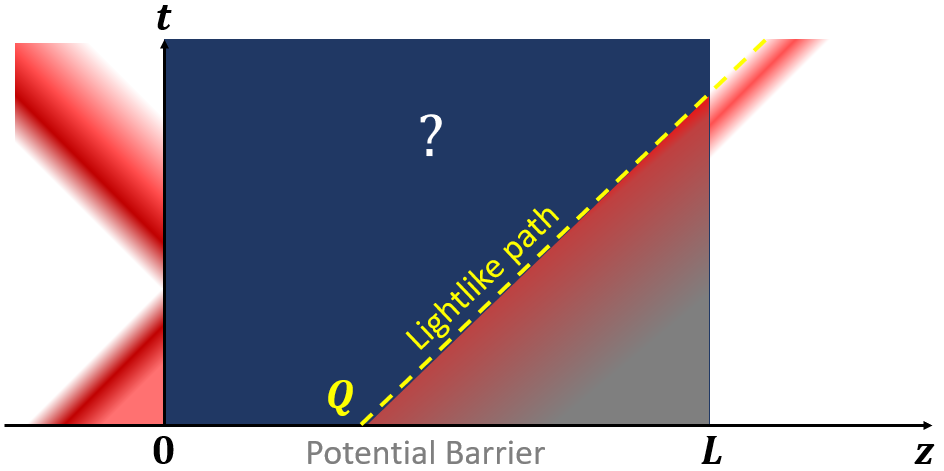}
	\caption{Visual representations of our arguments (i), (ii), and (iii), respectively, which are direct consequences of Theorem \ref{theo}. The shades of red represent the electron density $\Psi^\dagger \Psi$ (red large, white small). Upper panel: the field $\Psi$ is bound to propagate within the lightcone. Hence, electrons cannot travel faster than light. Quantum tunnelling through a barrier is no exception. Middle panel: Alice can generate a disturbance in $\Psi$ by altering the value of $A_\mu$ at her location. However, such disturbance travels slower than light, and it cannot reach Bob, who is causally disconnected from Alice (no superluminal signalling). Lower panel: the only way for a tunnelled wavepacket to exit at a time $t<L$ is that $\Psi \neq 0$ on the right of $Q$ already at $t=0$.}
	\label{fig:fig2}
	\end{center}
\end{figure}

Let us apply Theorem \ref{theo} to the relativistic quantum tunnelling, and let's see what we can argue on a purely mathematical basis.
\begin{itemize}
\item[(i)] One direct implication of Theorem \ref{theo} is that, if $\Psi=0$ in a region of space $\mathcal{R} \subset \Sigma$, then $\Psi=0$ also on $\mathcal{D}^+(\mathcal{R})$, the Cauchy development of $\mathcal{R}$ \cite{Wald}. This is a no-go theorem for superluminal motion: the electron cannot move faster than light, because the \textit{support} of the wavefunction cannot propagate outside the lightcone. For example, consider the situation illustrated in figure \ref{fig:fig2}, upper panel. A potential barrier extends over the region $0 \leq z \leq L$. At $t=0$, the electron is on the left of such barrier with probability 1, so that $\Psi=0$ for $z>0$.\footnote{Note that there is no obstruction to having wavefunctions that are of class $C^\infty$, and yet they vanish for $z>0$. The easiest way to construct them is to let $\Psi$ decay like $\sim \exp(1/z)$, as $z\rightarrow 0^-$.} Then, for arbitrary $t>0$, $\Psi$ must vanish in the region of space $z >t$. As a consequence, the probability for the electron to tunnel out of the barrier at a time $t<L$ is \textit{exactly} zero.
\item[(ii)] Theorem \ref{theo} enables us also to answer the most important question: can we use tunnelling electrons to send signals faster than light? In the literature, the word ``signal'' often generates some debate. However, when we have a partial differential equation like \eqref{diracequaz}, there is an unambiguous mathematical criterion to decide whether a superluminal signal can actually be sent or not. Consider the situation shown in figure \ref{fig:fig2}, middle panel. Alice and Bob are on the opposite sides of a potential barrier, and they are spacelike-separated. The electron wavefunction $\Psi$ fills the space between them. Can a decision of Alice affect a measurement of Bob? No! Alice can use a device to modify the value of the potential $A_\mu$ at her spacetime location. This indeed generates a perturbation in $\Psi$. However, from Theorem \ref{theo}, we know that changes in the value of $A_\mu$ outside the past lightcone of Bob cannot affect the value of $\Psi$ at Bob's spacetime location. Thus, Alice has no way to influence Bob's measurements\footnote{Another thing that Alice may do is to make a measurement herself. However, here Quantum Field Theory comes to our aid, reassuring us that spacelike-separated observables always commute, meaning that their measurements cannot influence each other \cite{Eberhard1989,Peskin_book,Coleman2018}.}.
\item[(iii)] When people say that ``the tunnelling effect is superluminal'', they typically have in mind the following scenario. A wavepacket meets a potential barrier; most of it is reflected, but a small portion tunnels through, and it appears on the other side earlier than a hypothetical light-beam emitted by the initial wavepacket (see figure \ref{fig:fig2}, lower panel). Recent findings already seem to question this picture \cite{Dumont2022}, but let us say (for the sake of argument) that the idea is somehow correct. What does Theorem \ref{theo} have to say about that? Suppose that $\Psi(t=0)$ were zero for $z>Q$ (figure \ref{fig:fig2}, lower panel). Then, by Theorem \ref{theo}, $\Psi(t)$ should vanish within the region $z>Q+t$, and there would be no tunnelled wavepacket. Therefore, the only way for us to observe a tunnelled wavepacket is to assume that $\Psi$ was already non-zero  on the right of $Q$ at $t=0$. In other words, to avoid a mathematical contradiction, we \textit{must} assume that the incoming wavepacket had a long tail, which extended largely inside the barrier, and that the tunnelled wavepacket is just the (subluminal) evolution of such  long tail. 
\end{itemize}

The conclusion of our point (iii) is very similar to that of \citet{Buttiker2003}: the tunnelled wavepacket is the causal evolution of the right tail of the incoming wavepacket, which enters the barrier much earlier than the peak, so that, if we only focus on the peaks, we get the illusion of a superluminal motion. \citet{Dumont2020} have criticised this interpretation, arguing that in quantum mechanics one should never say that one ``piece'' of the wavefunction originates from a corresponding ``piece'' in the past. Instead, the wavefunction should always be treated as whole. As a consequence, according to them, we cannot say that the tunnelled wavepacket ``originates'' from the right tail of the incoming wavepacket.

We do not wish to enter philosophical debates over the ontology of the wavefunction. On the other hand, we would like to point out that, when we say that the tunnelled wavepacket ``originates'' from the right tail, we are just making two rigorous mathematical statements (which follow from Theorem \ref{theo}). First, that if you change your initial data by removing the tail, i.e. by replacing $\Psi(t=0)$ with $\Psi(t=0)\Theta(Q-z)$, where $\Theta$ is the Heaviside step function\footnote{\label{foot}Readers might object that, by introducing the Heaviside function, we are no longer dealing with
continusouly differentiable functions, and thus Theorem \ref{theo} no longer applies.
But since the Dirac equation is a linear equation, Theorem \ref{theo} remains true
for distributional solutions (which will be the case for data involving the Heaviside function), see \cite{HormanderBook-AnalysisPD-2003-3}, Section 12.5. We assumed
continuous differentiability only in order to avoid technicalities and keep the proof short.}, the tunnelled wavepacket disappears. Second, if you instead  replace $\Psi(t=0)$ with $\Psi(t=0)\Theta(z-Q)$, leaving only the tail and cutting all the rest, the tunnelled wavepacket still remains, and it is completely unaffected. These facts may not establish an ``ontological relationship'' between the tunnelled wavepacket and tail of the incoming wavepacket, but they tell us that the \textit{existence} of the tunnelled wavepacket is a direct consequence of the \textit{existence} of such tail in its causal past. And this is enough to rule out any claim of superluminal behaviour. 

%\footnote{Although we would like to mention that, if we push the reasoning of \citet{Dumont2020} to the extremum, then any physical process involving quantum mechanics is superluminal. For example, consider a Gaussian wavepacket $\Psi \propto \exp(-z^2)$. Let it evolve, with arbitrary Hamiltonian, for a time $\Delta t=1$, and then perform a measurement, which collapses the particle at a given location, say $z=0$. Now, since $\Psi(z=10^{100})\neq 0$, we can apply the reasoning of \citet{Dumont2020}, and argue that the particle could have teleported itself from $z=10^{100}$ to $z=0$ in a time interval $\Delta t=1$.}

%We believe that this example clarifies the problem with the reasoning of \citet{Dumont2020}. When we assess the subluminality of a quantum-mechanical process, we must analyse how \textit{information} and \textit{probability} propagate. If we decide that a process is superluminal only because the electron could have come from any point within the support of the wavefunction, then superluminality becomes a tautological concept, since literally any wavefunction with non-compact support must exhibit superluminal effects.

\section{Subluminality in an inequality}\label{powerfuliequal}

\begin{figure}
\begin{center}
\includegraphics[width=0.5\textwidth]{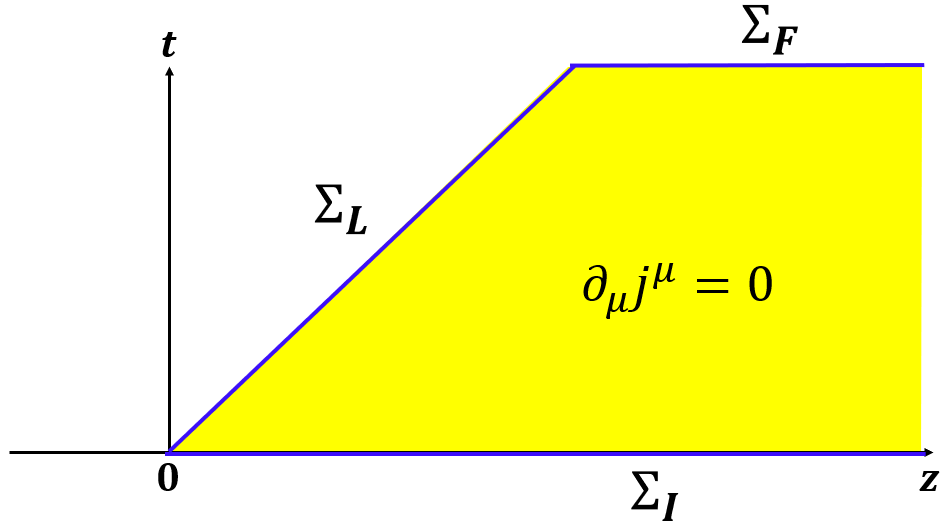}
	\caption{Visualization of the Gauss-theorem argument discussed in section \ref{powerfuliequal}. The yellow region represents the spacetime volume where we integrate the divergence of the Dirac current (which vanishes). The hypersurfaces $\Sigma_I$ and $\Sigma_F$ have constant time, and thus they are spacelike. The hypersurface $\Sigma_L$ is lightlike. Since the integral of $\Psi^\dagger \Psi$ across all space is normalised to $1$, we know that $\Psi$ decays to zero at spacelike infinity, and thus we can extend the integration region up to $z =+\infty$. Note that we have located the down-left corner of the trapezium in the origin only for convenience. This argument still holds if we translate the trapezium anywhere else.}
	\label{fig:fig3}
	\end{center}
\end{figure}

There is a simple inequality which, we believe, will convince even the most ardent ``superluminalist'' that quantum tunnelling is an entirely subluminal process. Let us consider the probability current $j^\mu = i\bar{\Psi} \gamma^\mu \Psi$, where $\bar{\Psi}=i \Psi^\dagger \gamma^0$. It can be easily shown that it has two properties  \cite{landau4,SakuraiAdvanced,dirac1981principles,
Finster1999,Finster2002}. First, it is conserved ($\partial_\mu j^\mu =0$), also in the presence of an external potential $A_\mu$. Second, it is non-spacelike, future-directed ($j^\mu j_\mu \leq 0$, $j^0 \geq 0$). In appendix \ref{AAA}, we verify explicitly that these properties hold also in the tunnelling model of \citet{Dumont2020}. Then, considering that $\Psi$ must decay to zero at spacelike infinity, we can apply the Gauss theorem over the (infinitely long) trapezoidal region shown in figure \ref{fig:fig3}, and we obtain (we adopt the orientation conventions of \citet{MTW_book}, section 5)
\begin{equation}\label{septe}
-\int_{\Sigma_I} j^\mu d \Sigma_\mu +\int_{\Sigma_L} j^\mu d \Sigma_\mu + \int_{\Sigma_F} j^\mu d \Sigma_\mu = 0 \, . 
\end{equation}
Since $j^\mu$ is non-spacelike future-directed, and $d\Sigma_\mu$, as a one-form, has positive sense towards the future (``standard orientation'' \cite{MTW_book}), then the integral over $\Sigma_L$ is non-negative, so that\footnote{This inequality is analogous to equation (10.1.11) of \citet{Wald}, which is used to prove well-posedness and causality of the Klein-Gordon equation. There, instead of $j^\mu$, \citet{Wald} uses an energy current, which is also conserved and future-directed non-spacelike. A similar theorem can also be found in \cite{GavassinoCausality2021}.}
\begin{equation}\label{laqualunque}
 \int_{\Sigma_F} j^\mu d \Sigma_\mu \leq \int_{\Sigma_I} j^\mu d \Sigma_\mu \, .
\end{equation}
On the other hand, on both $\Sigma_I$ and $\Sigma_F$ the integrand is just $j^0 d^3x$. But $j^0(x)=\Psi^\dagger(x) \Psi(x)$ is the probability density of observing the electron at $x$. Therefore, the inequality \eqref{laqualunque} reduces the following constraint:
\begin{equation}\label{inequalona}
\mathcal{P}_t(z>t) \leq \mathcal{P}_0(z>0) \, .
\end{equation}
In words: the probability of observing the electron on the right of $z=t$ at time $t$ will \textit{never} exceed the initial probability of observing that same electron on the right of $z=0$ at time $0$.

%Again, this means that the probability cannot ``flow'' outside the lightcone, and an electron can never ``overtake'' a photon that is moving in the same direction (even during a tunnelling process). In fact, if an electron reaches us before a photon, then it must have had an advantage over the photon, and the probability of such advantage \textit{cannot be less} than the probability of the electron arriving before the photon. 

\begin{figure}
\begin{center}
\includegraphics[width=0.6\textwidth]{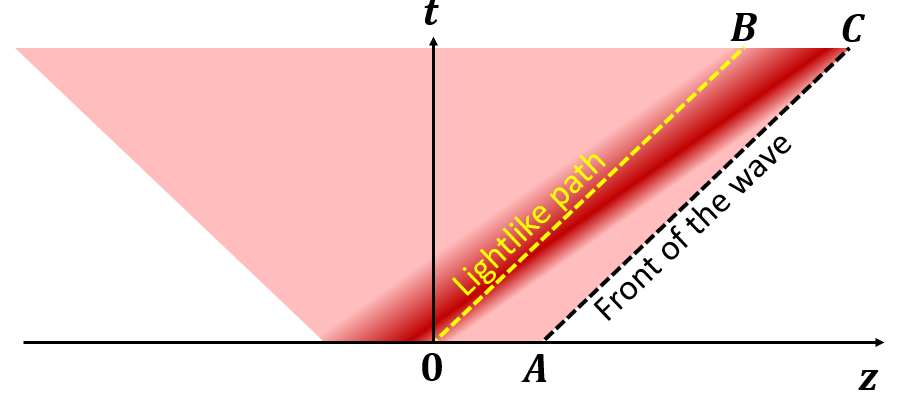}
	\caption{A forbidden process. A wavefunction has support over the pink region, which expands at the speed of light. Most of the probability density $j^0$ is initially located to the left of the origin (darker region). Can there be a large transfer of probability (dark red beam) from the left to the right of the yellow lightlike path? According to our inequality \eqref{inequalona}, no. In fact, the probability stored in the segment $BC$ can never exceed the probability stored in its non-zero causal past (the segment $0A$).}
	\label{fig:figImpossible}
	\end{center}
\end{figure}

To better understand the meaning of this result, consider the Minkowski diagram in figure \ref{fig:figImpossible}. The idea is the following. In Section III, implication (i), we proved that the front (i.e. the boundary) of the wavepacket cannot travel faster than light. However, one could argue that, even if the front itself is luminal, perhaps the main body of the wavefunction can still drift superluminally \textit{within} the support of the wavepacket, transiting from the left front to the right front faster than light, as in figure \ref{fig:figImpossible} (dark red beam). Our inequality \eqref{inequalona} forbids also this eventuality. In fact, a superluminal transfer of probability from the left to the right front would entail an increase of probability in a neighbourhood of the right front (i.e. on the segment $BC$ in figure \ref{fig:figImpossible}), and this would constitute a violation of \eqref{inequalona}. Note that a similar result would hold also in the absence of fronts (e.g. for Gaussian wavepackets): the probability on the right of $B$ at time $t(B)$ cannot be larger than the probability on the right of $0$ at time $t=0$.

Let us see other implications of equation \eqref{inequalona}.

\subsection{The tunnelling probability comes from the causal past of the tunnelled packet}

Let us apply equation \eqref{inequalona} to the setting discussed in point (iii) of section \ref{implicazia}. It is immediate to see that we can transport our Gauss-theorem argument of figure \ref{fig:fig3} into figure \ref{fig:fig2} (lower panel), locating the lower-left corner of the trapezium at $Q$ and overlapping $\Sigma_L$ with the lightlike path shown in figure \ref{fig:fig2}. Then, the inequality \eqref{inequalona} becomes
\begin{equation}\label{GloveofThanos}
\mathcal{P}(\text{``tunnelled packet''}) \leq \mathcal{P}(\text{``tail on the right of }Q\text{''}) \, .
\end{equation}
Again, this is showing us that the emerging wavepacket is just the subluminal evolution of the right tail of the incoming wavepacket. But now we know also something more: the probability associated to the tunnelled wavepacket cannot exceed that of this initial tail. This means that the tail cannot be used as ``springboard'', or a means to ``push'' the electron through the barrier faster than light. No way. The probability stored in the tail is the maximum probability that the tunnelled wavepacket can carry. 

There is a subtlety that we need to mention. In figure \ref{fig:fig2} (lower panel), the point $Q$ falls inside the barrier (i.e. $Q>0$). But this is true only if we set our clocks in such a way that at $t=0$ the incoming wavepacket is about to enter the barrier. In numerical experiments like the one performed by \citet{Dumont2020}, the wavepacket is on the far left of the barrier at $t=0$. In this case, equation \eqref{GloveofThanos} still holds, but the point $Q$ will also be on the far left of the barrier ($Q \ll 0$). In Appendix \ref{BBBBBBBBBBBBB}, we calculate the position of $Q$ for the numerical experiment of  \citet{Dumont2020}, and we verify explicitly that their numerical analysis corroborates equation \eqref{GloveofThanos}. 

\subsection{Luminal bound on the speed of tunnelling}\label{luminalbound}

We are finally able to prove that quantum tunnelling is not instantaneous, and that its speed is bounded above by the speed of light. Our proof works as follows.

Suppose that, at time $t=0$, the electron is on the left of the barrier, and it is about to tunnel through. The left edge of the barrier is located at $z=0$ (as usual), while the right edge is at $z=L$. To keep our discussion completely general, we allow for a little portion of the electron's wavefunction to have already ``leaked'' inside the barrier. Hence, we just assume that the electron is on the left of the barrier with some probability $\mathcal{P}\leq 1$. Thus, we have the initial condition $\mathcal{P}_0(z\leq 0)=\mathcal{P}$, where the subscript ``$0$'' means that we are evaluating the probability at $t=0$. Since the total probability is normalised to $1$, we also know that $
   \mathcal{P}_0(z>0) = 1-\mathcal{P}=\text{``Leaked probability''}
$.
Plugging this initial condition into equation \eqref{inequalona}, we find that 
\begin{equation}\label{fbuf}
1-\mathcal{P} \geq \mathcal{P}_t(z>t) \, .    
\end{equation}
Now, let us set $t < L$. Then, the probability $\mathcal{P}_t(z>t)$ cannot be smaller than the probability $\mathcal{P}_t(z>L)$:
\begin{equation}\label{slarb}
\mathcal{P}_t(z>t)= \mathcal{P}_t(L>z>t)+\mathcal{P}_t(z>L) \geq \mathcal{P}_t(z>L).    
\end{equation}
Combining \eqref{fbuf} and \eqref{slarb}, we arrive at the inequality  $1-\mathcal{P} \geq \mathcal{P}_t(z>L)$, for $t<L$. On the other hand, $ \mathcal{P}_t(z>L)$ is the probability of detecting the electron on the right of the barrier at time $t$. But this is just the tunnelling probability at time $t$. In conclusion, we have that
\begin{equation}
    \mathcal{P}_t (\text{``Tunnelling''}) \leq \mathcal{P}_0(\text{``Leaked''}) \, , \spc \text{for }t<L\, .
\end{equation}
In a nutshell, this is telling us that the best we can get in a time $t<L$ is that the part of the wavefunction that is already inside the barrier (at $t=0$) will emerge on the right. But nothing more than this. If this ``leaked tail'' is negligible, then the tunnelled wavepacket cannot emerge in a time smaller than $L/c$ (restoring non-geometric units).

\section{Subluminality as an algebraic identity}

All our analysis till this point has been carried out with explicit reference to the wavefunction $\Psi$. It is natural to wonder whether we can also express our results using Dirac's ``bra-ket notation''. This is what we aim to do here. For clarity, we switch to $1+1$ dimensions, and we adopt rectangular coordinates $(t,z)$.

Since in the bra-ket notation one only deals with quantum states $\ket{\Psi}$, with no explicit reference to spacetime events and locations, we need to first express Theorem \ref{theo} in a slightly different way. Our reasoning is the following. Since the Dirac equation is linear, we can always express a solution $\Psi$ as the superposition of two other solutions, $\Psi=\Psi_L+\Psi_R$, provided that the initial data for $\Psi_L$ and $\Psi_R$ add up to the initial data of $\Psi$, namely $\Psi(0,z)=\Psi_L(0,z)+\Psi_R(0,z)$. We choose for these two solutions the following initial data: 
$
\Psi_L(0,z)=\Psi(0,z)\Theta(-z)$, and $\Psi_R(0,z)=\Psi(0,z)\Theta(z)$, which clearly add up to $\Psi(0,z)$. On the other hand, our theorem (which holds also for distributional solutions, see footnote \ref{foot}) guarantees that, since $\Psi_L(0,z)=0$ for $z>0$, then $\Psi_L(t,z)=0$ for $z>t$. This implies that
\begin{equation}\label{zumba}
\Psi(t,z)=\Psi_R(t,z) \quad \text{for }z>t.
\end{equation}
This clearly shows that the ``part of the wavefunction'' that is initially (at $t=0$) in $z<0$ cannot travel into the region $z>t$. Equivalently, the part of the wavefunction that enters the region $z>t$ is the causal evolution of $\Psi_R(0,z)$, namely the portion of the initial wavefunction that was in the causal past of the region $z>t$. 

Let us now switch to bra-ket notation. If we work in the Schr\"{o}dinger picture, the initial wavefunctions $\Psi(0,z)$, $\Psi_L(0,z)$, $\Psi_R(0,z)$ correspond to three different quantum states, which may be represented by three corresponding state vectors: $\ket{\Psi}$, $\ket{L}$, and $\ket{R}$. The first state vector is normalised, $\braket{\Psi}{\Psi}=1$, while
\begin{equation}\label{sixteen}
\begin{split}
\braket{L}{L} ={}& \int_{\mathbb{R}} \Psi^\dagger(0,z)\Psi(0,z)\Theta(-z)dz = \mathcal{P}_0(z<0) \, , \\
\braket{R}{R} ={}& \int_{\mathbb{R}} \Psi^\dagger(0,z)\Psi(0,z)\Theta(z)dz = \mathcal{P}_0(z>0) \, , \\
\braket{L}{R} ={}& \int_{\mathbb{R}} \Psi^\dagger(0,z)\Psi(0,z)\Theta(z)\Theta(-z) dz =0 \, . \\
\end{split}
\end{equation}
Clearly, the initial condition $\Psi(0,z)=\Psi_L(0,z)+\Psi_R(0,z)$ is equivalent to $\ket{\Psi}=\ket{L}+\ket{R}$. Then, the condition $\Psi=\Psi_L+\Psi_R$ just expresses the fact that unitary time evolution (in the Schr\"{o}dinger picture) is linear:
\begin{equation}\label{gringoNN}
e^{-i\hat{H}t}\ket{\Psi}= e^{-i\hat{H}t}\ket{L}+e^{-i\hat{H}t}\ket{R} \, ,
\end{equation}
where $\hat{H}$ is the Hamiltonian which generates the dynamics of \eqref{diracequaz}.
Finally, the fact that $\Psi_L(t,z)=0$ for $z>t$ translates into the condition $\hat{\mathcal{P}}(z>t)e^{-i\hat{H}t}\ket{L}=0$, where $\hat{\mathcal{P}}(z>t)$ is the orthogonal projector\footnote{The orthogonal projector $\hat{\mathcal{P}}(z>Q)$ maps an arbitrary wavefunction $\Psi$ into $\Theta(z-Q)\Psi$, while the orthogonal projector $\hat{\mathcal{P}}(z<Q)$ maps an arbitrary wavefunction $\Psi$ into $\Theta(Q-z)\Psi$.} onto the region $z>t$. Therefore, if we apply $\hat{\mathcal{P}}(z>t)$ on both sides of \eqref{gringoNN} we recover equation \eqref{zumba}, namely
\begin{equation}\label{pzt}
\hat{\mathcal{P}}(z>t)e^{-i\hat{H}t}\ket{\Psi}= \hat{\mathcal{P}}(z>t)e^{-i\hat{H}t}\ket{R} \, .
\end{equation}
Equation \eqref{pzt} expresses the subluminality of the Dirac equation, both in the presence and in the absence of a potential barrier. In fact, it tells us that, if we perform a measurement inside the region $z>t$, there are no contributions coming from $\ket{L}$. In other words, all probabilities computed inside the region $z>t$ depend only on the state $\ket{R}$, which describes that part of the wavefunction which was in the causal past of such region. Again, this shows that relativistic quantum tunneling is an entirely subluminal process. This conclusion is also corroborated by the observations below:
\begin{itemize}
\item If we take the norm of equation \eqref{pzt}, we obtain
\begin{equation}
\bra{\Psi}e^{i\hat{H}t}\hat{\mathcal{P}}(z>t)e^{-i\hat{H}t}\ket{\Psi} = \bra{R}e^{i\hat{H}t}\hat{\mathcal{P}}(z>t)e^{-i\hat{H}t}\ket{R} \leq \bra{R}e^{i\hat{H}t}e^{-i\hat{H}t}\ket{R}=\braket{R} \, .
\end{equation}
On the other hand, the quantity $\bra{\Psi}e^{i\hat{H}t}\hat{\mathcal{P}}(z>t)e^{-i\hat{H}t}\ket{\Psi}$ is just the probability of observing the electron in the region $z>t$ at time $t$, namely $\mathcal{P}_t(z>t)$. Hence, recalling the second equation of \eqref{sixteen}, we recover the inequality \eqref{inequalona}. The interpretation is simple: the probability stored in $z>t$ at time $t$ comes only from $\ket{R}$, and hence it cannot exceed $\braket{R}=\mathcal{P}_0(z>0)$.
\item It is evident that $\ket{L}=\hat{\mathcal{P}}(z<0)\ket{\Psi}$. Hence, the condition $\hat{\mathcal{P}}(z>t)e^{-i\hat{H}t}\ket{L}=0$, which is valid for any $\ket{\Psi}$, can be expressed as an operatorial identity:
\begin{equation}
\hat{\mathcal{P}}(z>t)e^{-i\hat{H}t}\hat{\mathcal{P}}(z<0) =0 \, .
\end{equation}
In words: No contribution coming from $z<0$ can reach the region $z>t$ in a time $t$, independently from the presence of a potential barrier.
\end{itemize}
Let us make one final remark. In the analysis above, we have expressed the initial state $\ket{\Psi}$ as the quantum superposition of two other states, $\ket{L}$ and $\ket{R}$. \citet{Dumont2020} have criticised this kind of approach. Here we report their reasoning \cite{Dumont2020}:
``This argument appears to imply that we are actually able to track individual components of the initial
wavepacket (...), and identify parts of the final wavepacket as having come from
the front or back of the initial distribution (...). This implication is problematic, as it
would only be possible to track the individual cars in this way with access to some hidden variables as in the
case of Bohmian mechanics and the like.''

In response to this, we would like to remark that the principle of superposition is a defining feature of quantum mechanics, and the ability to track individual parts of a wavepacket follows directly from the linearity property of the unitary evolution, see equation \eqref{gringoNN}. Equation \eqref{pzt}, then, only expresses the fact that the outcome of a measurement performed inside the spacetime region $z>t$ depends only on $\ket{R}$, which is the projection of $\ket{\Psi}$ into the causal past of such region. This is an observable physical fact. No reference to interpretations of quantum mechanics is needed. 

\section{Conclusions}

 General Relativity \cite{carroll_2019,Wald,Hawking1973} and the theory of partial differential equations \cite{Rauch_book,CourantHilbert2_book} provide us with all the machinery necessary to assess the speed of a physical process, and the propagation of information within a given theory. Here, we have applied these techniques to relativistic quantum tunnelling, modelled using the classical Dirac equation in a background electromagnetic field. This has allowed us to establish rigorously three mathematical facts.
\begin{itemize}
    \item It impossible to use quantum tunnelling to send information faster than light [see Section \ref{implicazia}, point (ii)]. In fact, if an observer (Alice) perturbs the wavefunction $\Psi$ at a point, the perturbation is bound to travel inside the lightcone. This implies that, if another observer (Bob) sits somewhere outside the lightcone, he cannot know that Alice has acted on the electron, because the perturbation cannot reach him. Bob has no way to tell whether Alice perturbed the wavefunction or not. We would like to stress that this result is completely independent from any bound on the ``speed of tunnelling''. In fact, this theorem just tells us that, if quantum tunnelling were superluminal, Alice would have absolutely no influence on that part of the wavefunction that exits her future lightcone. She would not be able to control it in any way. It would be impossible for her to manipulate its shape, or even to stop it. The very fact that a ``superluminal wavepacket'' reaches Bob would be independent from Alice's decisions. So, even if quantum tunelling were superluminal, it would be impossible to use it to send \textit{information}, because Bob would have no way to infer what Alice has done.
    \item  Relativistic quantum tunnelling is not ``instantaneous''. Instead, its speed is bounded  by the speed of light. In particular, if at a given time the electron is on the left of the barrier with probability $\mathcal{P}$, then we need to wait at least a time $L/c$ (``length of barrier''/``speed of light''), before the probability of having the electron on the right of the barrier can become larger than $1-\mathcal{P}$ [see Section \eqref{luminalbound}]. This guarantees that the probability ``flows'' subluminally between the two edges of the barrier. As a particular case, if the electron is on the left of the barrier with probability $1$, we need to wait a time $L/c$ before it can emerge on the right with non-zero probability [see Section \ref{implicazia}, point (i)].
    \item If a photon is travelling towards the right, the probability of observing an electron on the right of such photon can only decrease in time (or stay constant). In other words, photons always ``overtake'' electrons. The reversal cannot happen, even during quantum tunnelling. This is shown in figure \ref{fig:figImpossible}.
\end{itemize}

Hence, we believe that this article has finally settled a 20-year-old debate. The Dirac equation is a perfectly causal field equation, also when we turn on an extremely high potential barrier. That is because the electromagnetic four-potential $A_\mu (x)$ does not enter the equation of the characteristics. Indeed, recent numerical tests \cite{Dumont2022} confirm our main message: tunnelling electrons cannot be faster than photons in vacuum. Interestingly, if we assume that a photon and an electron start with the same initial distribution, then our inequality \eqref{inequalona} can be equivalently rewritten in the form\footnote{To show this, just consider one electron and one photon with same initial probability distributions, and use the fact that, for photons, the inequality \eqref{inequalona} is saturated.}
\begin{equation}
\mathcal{P}_t(\text{``Arrival electron''}) \leq \mathcal{P}_t (\text{``Arrival photon''}) \, ,
\end{equation}  
which is precisely what has been observed in all tests performed by \citet{Dumont2022}.

We hope that our work will also foster the interdisciplinary communication between the quantum physics community and the mathematical relativity community.

\section*{Acknowledgements}

M.M.D. is partially supported by a Sloan Research Fellowship provided by the Alfred P. Sloan foundation, NSF grant DMS-2107701, and a Vanderbilt's Seeding Success Grant.
L.G. is partially supported by a Vanderbilt's Seeding Success Grant.
The authors would like to thank the anonymous referees for several comments
that lead to a significant improvement of the manuscript over its 
first version.

\appendix

\section{Probability current for tunnelling models}\label{AAA}

The tunnelling model of \citet{Dumont2020} is (1+1)-dimensional, and it evolves only two components of $\Psi$ in the Dirac basis (as the other two components are fully decoupled). We call such components ``$\, f \,$'' and ``$\, h \,$''. The Dirac equation then reads
\begin{equation}\label{DiracOraEpersembre1d}
\left\{ 
\begin{array}{ll} 
i\partial_t f =& i\partial_z h+ (V+m)f \, ,\\
i\partial_t h =& i\partial_z f+ (V-m)h \, ,\\
\end{array}\right.
\end{equation} 
where $V=V(x)$ is the potential barrier.
Taking the complex conjugate, we get
\begin{equation}
\left\{ 
\begin{array}{ll} 
i\partial_t f^* =& i\partial_z h^* - (V+m)f^* \, ,\\
i\partial_t h^* =& i\partial_z f^* - (V-m)h^* \, .\\
\end{array}\right.
\end{equation}
Thus, it is immediate to verify that
\begin{equation}
\left\{ 
\begin{array}{ll} 
\partial_t (f^* f) =& f^* \partial_z h + f \partial_z h^* \, ,\\
\partial_t (h^* h) =& h^* \partial_z f + h \partial_z f^* \, .\\
\end{array}\right.
\end{equation}
As we can see, all the terms with $V$ cancel out.
Taking the sum of these two equations, and bringing every term on the left-hand side, we obtain an equation of the form $\partial_\mu j^\mu =0$, with
\begin{equation}\label{lacurrent}
j^\mu = 
\begin{pmatrix}
f^* f + h^* h \\
-f^* h - h^* f \\
\end{pmatrix} \, .
\end{equation}
As we can see, $j^0 = f^* f + h^* h$ is non-negative definite, and it has the usual form of a probability density. To prove that $j^\mu$ is non-spacelike future-directed, we only need to show that $j^0 \geq |j^z|$, namely $f^* f + h^* h \geq |f^* h + h^* f |$. But this follows immediately from the chain of identities below:
\begin{equation}
0 \leq |f \pm h|^2 = (f^* \pm h^*) (f \pm h)= f^* f + h^* h \pm (f^* h + h^* f ) \, .
\end{equation}

\subsection{Application: superluminal interference fringes do not transport probability}

Some solutions of the Dirac equation can exhibit interference fringes whose phase velocity is larger than the speed of light \cite{Berry2012}. We know, from Theorem \ref{theo}, that such fringes cannot be used to transport information faster than light. Now we are also in the position to show that they cannot even be used to transport the electron itself (namely, the probability density) faster than light.
The proof is very simple. We have shown that the probability satisfies a continuity equation of the form $\partial_t j^0+\partial_z j^z=0$ (in 1+1 dimensions, for simplicity). If we define the velocity $v:=j^z/j^0$ (``probability flux''/``probability density''), the continuity equation acquires the usual form that we meet in hydrodynamics textbooks \cite{landau6}: $\partial_t j^0+\partial_z(j^0 v)=0$. This justifies the interpretation of $v$ as the ``velocity of probability'', since it quantifies how fast a given probability density is crossing the boundary of a certain region of space:
\begin{equation}
\dfrac{d}{dt} \mathcal{P}_t(z<a)=\dfrac{d}{dt} \int_{-\infty}^a j^0(t,z) dz = -j^0(t,a) v(t,a) \, .
\end{equation} 
On the other hand, we have shown that $|j^z| \leq j^0$, which implies $|v|\leq 1$. Hence, the speed of probability can never exceed the speed of light. This completes our proof.

It is instructive to analyse a concrete example. To simplify the calculations, let us set $m=V=0$ in equation \eqref{DiracOraEpersembre1d}. Then, it is immediate to see that
\begin{equation}\label{DiracOraEpersembregab}
\left\{ 
\begin{array}{ll} 
f(t,z)=a(z+t)+b(z-t) \, ,\\
h(t,z)=a(z+t)-b(z-t) \, ,\\
\end{array}\right.
\end{equation}
is a solution of the massless Dirac equation, for any couple of complex functions $a$ and $b$. Equation \eqref{DiracOraEpersembregab} describes a quantum superposition of a left-travelling state $a$ and a right-travelling state $b$. It is immediate to verify that
\begin{equation}\label{lequattrolinee}
\begin{split}
f^* f ={}& a^* a +a^*b +b^* a + b^* b \, , \\
h^* h ={}& a^* a -a^*b -b^* a + b^* b \, ,\\
f^* h ={}& a^* a -a^*b +b^* a - b^* b \, ,\\
h^* f ={}& a^* a +a^*b -b^* a - b^* b \, ,\\
\end{split}
\end{equation}
where it is understood that all functions $a$ are evaluated at $z+t$, and all functions $b$ are evaluated at $z-t$. If we plug the above formulas into \eqref{lacurrent}, we finally obtain a formula for the velocity of probability:
\begin{equation}
v = \dfrac{b^* b-a^* a}{b^* b+a^* a} \in [-1,1] \, .
\end{equation}
Now, let us set $a(z)=\exp(ikz)$ and $b(z)=\exp(ipz)$, with $p>k>0$. Then, the first line of \eqref{lequattrolinee} becomes
\begin{equation}
f^* f = 4 \cos^2 \bigg[ \dfrac{p-k}{2} \bigg( z- \dfrac{p+k}{p-k} \, t \bigg) \bigg] \, .
\end{equation}
As we can see, the wavefunction exhibits an interference fringe, which drifts with phase velocity $(p+k)/(p-k)>1$. On the other hand, since $a^* a=b^* b=1$, the probability has vanishing net velocity, i.e. $v=0$. This result is completely analogous to what we see in hydrodynamics: a sound wave travels at the speed of sound, but this does not mean that matter itself is transported at the speed of
sound. Indeed, during the passage of a sound wave, the matter elements oscillate, but on average they do not move. It is the same here.
The phase velocity of an interference fringe can exceed the speed of light, but such fringe does not transport probability. The particle is, on average, at rest.

\section{A numerical experiment}\label{BBBBBBBBBBBBB}

Here, we show that the numerical experiment of \citet{Dumont2020} corroborates our equation \eqref{GloveofThanos}. We choose our units of space, time and energy in such a way that $c=\hbar=\lambda_{rc}=1$, where $\lambda_{rc}$ is the reduced Compton wavelength of the electron. Furthermore, we set the origin of the coordinate $z$ to coincide with the left edge of the barrier.

\begin{figure}
\begin{center}
\includegraphics[width=0.4\textwidth]{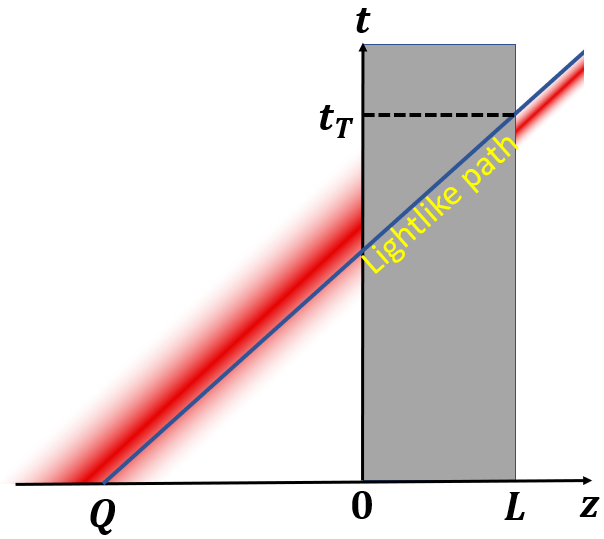}
	\caption{Location of the point $Q$ in the numerical experiment of \citet{Dumont2020}. The mathematical procedure for computing it is the following. First step: identify the position of the tunnelled wavepacket in the Minkowski diagram (red ``beam'' in the up-right corner). Second step: draw a right-travelling lightlike path (in blue) located immediately on the left of the tunelled wavepacket. Third step: find the location $Q$ where this lightlike path intersects the line $t=0$. The region $z>Q$ defines the ``tail'' of the incoming wavefunction in the causal past of the tunnelled wavepacket. Equation \eqref{GloveofThanos} states that the tunnelling probability cannot exceed the probability associated to such tail (because the tunnelled wavepacket is the causal evolution of such tail). In other words, the probability flows \textit{subluminally} from the tail to the tunnelled wavepacket.  We note that, since the incoming wavepacket is very far from the barrier at $t=0$, the point $Q$ that marks the beginning of the tail lies outside the barrier ($Q<0$).}
	\label{fig:BBB}
	\end{center}
\end{figure}

The incoming wavepacket is initially centered around $z_0 =-120$. The barrier starts at $z=0$ and ends at $z=L=15$ (we consider the ``most superluminal'' case). The tunnelled wavepacket emerges at a time $t_T \approx 110$ (see figure 1.c of \cite{Dumont2020}). Therefore, the point $Q$ is just (see figure \ref{fig:BBB})
\begin{equation}
Q=L-t_T=-95 \, .
\end{equation} 
This is telling us that the tail of the wavefunction that lies inside the causal past of the tunelled wavepacket covers the region $z>-95$. If our equation \eqref{GloveofThanos} is correct, the probability associated to such tail should not be smaller than the probability associated to the tunnelled wavepacket. Now, the initial probability distribution is essentially a Gaussian,
\begin{equation}
j^0(0,z) \approx \dfrac{1}{\sqrt{\pi \Delta z^2}} \exp \bigg[ -\dfrac{(z-z_0)^2}{\Delta z^2} \bigg] \, ,
\end{equation}
with width $\Delta z = 15$. Integrating it for $z>Q$, we get
\begin{equation}
\mathcal{P}_0(z>-95) \approx 0.009 \, .
\end{equation}
This is a small tail, as it encompasses only $\sim 1\%$ of the incoming wavepacket, but it is much larger than $\mathcal{P}(\text{``tunnelled packet''}) \sim 10^{-15}$ (see figure 1.c of \citet{Dumont2020}). This corroborates equation \eqref{GloveofThanos}, and it confirms that no superluminal ``flow'' of probability has taken place. The probability associated to the tunnelled wavepacket never exceeds the probability that was already contained in its causal past.

\bibliography{Biblio}

\label{lastpage}

\end{document}